\documentclass[aip,pop,amsfonts, floatfix, amssymb, amsmath ,reprint ,subscriptadress]{revtex4-1}

\usepackage{breqn}
\usepackage{graphicx}
\usepackage{dcolumn}   
\usepackage{bm}        
\usepackage{amssymb}   
\usepackage{soul}
\usepackage{gensymb}
\usepackage[normalem]{ulem}

\usepackage[breaklinks=true,colorlinks=true,linkcolor=blue,urlcolor=blue,citecolor=blue]{hyperref}

\usepackage{float}

\begin{document}

\title{Assessing global ion thermal confinement in critical-gradient-optimized stellarators}

\author{A.~Ba\~n\'on~Navarro}
\email{alejandro.banon.navarro@ipp.mpg.de}
\affiliation{Max Planck Institute for Plasma Physics Boltzmannstr 2 85748 Garching Germany}
\author{G.~T.~Roberg-Clark} 
\affiliation{Max Planck Institute for Plasma Physics Wendelsteinstr 1 17491 Greifswald Germany}
\author{G.~G.~Plunk} 
\affiliation{Max Planck Institute for Plasma Physics Wendelsteinstr 1 17491 Greifswald Germany}
\author{D.~Fernando} 
\affiliation{Max Planck Institute for Plasma Physics Boltzmannstr 2 85748 Garching Germany}
\author{A.~Di~Siena} 
\affiliation{Max Planck Institute for Plasma Physics Boltzmannstr 2 85748 Garching Germany}
\author{F.~Wilms} 
\affiliation{Max Planck Institute for Plasma Physics Boltzmannstr 2 85748 Garching Germany}
\author{F.~Jenko} 
\affiliation{Max Planck Institute for Plasma Physics Boltzmannstr 2 85748 Garching Germany}

\newcommand{\bcomment}[1]{\textcolor{blue}{\bf #1}}

\begin{abstract}
We investigate the confinement properties of two recently devised quasi-helically symmetric stellarator configurations, HSK and QSTK. Both have been optimized for large critical gradients of the ion temperature gradient mode, which is an important driver of turbulent transport in magnetic confinement fusion devices. To predict the resulting core plasma profiles, we utilize an advanced theoretical framework based on the gyrokinetic codes GENE and GENE-3D, coupled to the transport code TANGO. Compared to the HSX stellarator, both HSK and QSTK achieve significantly higher core-to-edge temperature ratios, partly thanks to their smaller aspect ratios, with the other part due to more detailed shaping of the magnetic geometry achieved during optimization.  The computed confinement time, however, is less sensitive to core temperature than edge temperature, simply due to the disproportionate influence the edge has on stored plasma energy. We therefore emphasize the possible benefits of further optimizing turbulence in the outer core region, and the need to include accurate modelling of confinement in the edge region in order to assess overall plasma performance of turbulence optimized stellarators.
\end{abstract}

\maketitle


\begin{section}{Introduction}

In present-day fusion research, stellarators are widely recognized as an attractive alternative to tokamaks. Their distinct advantages include steady-state operation, the absence of disruptions, and no dependency on external current drive systems~[\onlinecite{Helander_2012}]. Although stellarators need to be carefully optimized with respect to a wide range of sometimes conflicting physics and engineering criteria, the large number of plasma shaping parameters may enable the identification of new and promising designs on the path to future fusion power plants.

A key aspect of this challenge has to do with the optimization of the magnetic confinement properties~[\onlinecite{Pedersen_2018, wolf_2019, Klinger_2019, Bozhenkov_2020}]. While the neoclassical transport in modern stellarators like W7-X has been significantly reduced, the turbulent transport has emerged as a limiting factor for plasma performance. This motivates the development of techniques for designing turbulence optimized stellarators.  In particular the idea of ``critical gradient'' optimization has been developed in several papers in recent years [\onlinecite{clark_2022, robergclark2022reduction, qstk}], resulting in two configurations studied here.  Other approaches have more recently begun to emerge ~[\onlinecite{edi2023,rogerio_2023}].  Global transport studies, for the first time done here, are essential to determine how such optimization translates into temperature profiles and overall plasma performance.

In the following, we will employ an extended version of an advanced theoretical framework which was used successfully to describe the confinement degradation (ion temperature clamping) observed in the first experimental campaign of W7-X, as evidenced by the W7-X electron-heated experiments~[\onlinecite{Beurskens_2021}, \onlinecite{Carralero_2021}]. In our previous work~[\onlinecite{Navarro_NF_2023}], we examined this phenomenon using a novel integrated modeling tool designed specifically for modern-optimized stellarator devices, which self-consistently incorporates turbulence, neoclassical transport, the radial electric field, and 3D effects.  It involves the coupling of the global gyrokinetic turbulence code GENE-3D~[\onlinecite{Maurer2020}, \onlinecite{wilms_2021}], the neoclassical transport code KNOSOS~[\onlinecite{velasco_2020}, \onlinecite{velasco_2021}], and the 1D transport code TANGO~[\onlinecite{Parker_2018}, \onlinecite{Di_Siena_2022}]. By coupling these codes, which takes advantage of the time scale separation between turbulence and transport phenomena, we achieved a considerable reduction in computational cost. Using this framework, we successfully simulated the evolution of ion temperature profiles for electron-heated plasmas in W7-X, thereby reproducing the ion temperature clamping observed in the experiments.

In the present paper, we improve on the previous work by introducing an additional intermediate layer of fidelity that provides the flexibility of using the local (flux-tube) gyrokinetic code GENE, instead of just the global gyrokinetic code GENE-3D, expediting the calculation of the steady-state profiles. Using this updated framework, we conduct an investigation of the confinement properties of HSK (Helically-Symmetric Kompakt stellarator)~[\onlinecite{robergclark2022reduction}] and QSTK (Quasi-Symmetric Turbulence Konzept)~[\onlinecite{qstk}], two recently found quasi-helical stellarators that have been optimized for a large critical gradient of the ion temperature gradient (ITG) mode. HSK boasts exceptional quasi-symmetry, low neoclassical transport, and minimal alpha particle losses, although at the expense of MHD instability. QSTK, on the other hand, combines those attractive attributes with MHD stability and an acceptable level of coil complexity. We compare the performance of those devices to that of HSX (Helically Symmetric eXperiment)~[\onlinecite{hsx}] at the University of Wisconsin-Madison. Our simulations examine the behavior of both stellarators when exposed to external heating sources with varying power levels. In particular, we investigate the correlation between the ion temperature profile, the normalized ion temperature gradient, and the heating power. Additionally, we compare the resulting energy confinement times in the context of an ISS04-like scaling law derived from the simulation results.

The remainder of the present paper is organized as follows. In Section~\ref{optimization}, some general considerations regarding turbulence optimized stellarators are presented, to provide some context for the present study.
In Sections~\ref{Model} and~\ref{setup}, we introduce, respectively, the model description and the simulation setup. The simulation results are presented and discussed in Section~\ref{results}, and a summary of the main findings as well as some general conclusions derived from our work are presented in Section~\ref{conclusions}.

\end{section}


\begin{section} {Turbulence Optimized Stellarators  \label{optimization}}

Before we focus on the more technical aspects of our study, we would like to put forth some general considerations regarding turbulence optimized stellarators, to provide some context.

Intense experimental and theoretical research dedicated to tokamaks over many years has firmly established that turbulent transport can be influenced by a number of control parameters. Among the latter are various properties of the magnetic geometry~[\onlinecite{rice_1996}, \onlinecite{austin_2019}] as well as the presence of radial electric fields~[\onlinecite{burrel_2020}] and/or energetic particles~[\onlinecite{Citrin_2023}]. In addition, many nonlinear feedback loops between the plasma temperature/density profiles and the turbulent energy/particle fluxes have been discovered~[\onlinecite{Angioni_2021}, \onlinecite{siena_2021}]. This represents a wealth of knowledge that can be exploited in the development of techniques for the turbulence optimization of stellarators.

As is well established, the existence of turbulent transport in fusion devices has a range of implications. Some negative impacts include a low normalized temperature gradient (as in the aforementioned ``temperature clamping") across the plasma volume, as well as a reduction of the energy confinement time via significant transport, both of which can significantly impact the fusion triple product. On the positive side, turbulence may facilitate the transport of plasma impurities to avoid accumulation of helium ash and wall materials. In this context, the plasma edge region constitutes a further challenge, given the fact that edge turbulence controls the heat and particle exhaust as well as the plasma wall interaction while also exerting a large influence on the overall magnetic confinement properties by setting the boundary conditions for the core plasma profiles.

Although many related open questions exist, a successful attempt at optimizing the core ion temperature profiles has recently been made, with the creation of two newly designed quasi-helical stellarators that have been optimized for large critical gradients of the ion temperature gradient (ITG) mode, HSK~[\onlinecite{robergclark2022reduction}] and QSTK~[\onlinecite{qstk}].  The present work quantitatively addresses how critical gradient optimization affects temperature profiles and ion energy confinement time.  The existing HSX stellarator, which is not turbulence optimized, will serve as a point of comparison.

We carry out first-principles-based (gyrokinetic) simulations of ITG turbulence in these three devices, computing the core-to-edge ratio of the ion temperature profile and the energy confinement time -- assuming a fixed edge temperature. Although the employed GENE and GENE-3D codes are able to include kinetic electrons, multiple ion species, magnetic field fluctuations, and collisions, all of these effects will be neglected in the present study, to stay in line with the simulation setups used in the studies leading to the design of HSK and QSTK. We note that, as a consequence of these simplifications, electron heat transport and the transport of particles and impurities are not evaluated.

\end{section}


\begin{section} {Model description \label{Model}}

The present study utilizes a state-of-the-art theoretical framework to predict plasma core profiles for stellarators. The framework is an extension of the approach employing the global gyrokinetic stellarator turbulence code GENE-3D, the neoclassical code KNOSOS, and the transport code TANGO, as introduced in Ref.~[\onlinecite{Navarro_NF_2023}]. To accelerate the convergence of the solution, an additional intermediate layer of fidelity has been incorporated, namely the option of also using the local flux-tube gyrokinetic code GENE~[\onlinecite{Jenko_PoP2000}]. The methodology involves starting with the lower fidelity, faster GENE model and subsequently refining the solution using GENE-3D. If the flux-tube solution is already in close proximity, this will significantly expedite the convergence of the more costly GENE-3D simulations. A schematic diagram of the workflow is presented in Figure~\ref{fig:fig0}. First, KNOSOS computes the neoclassical fluxes and radial electric field for a specific magnetic equilibrium and density/temperature profiles. Then, GENE uses these plasma profiles and shear of the radial electric field to evaluate the turbulent transport.  Finally, the TANGO code is used to evaluate new plasma profiles that are consistent with both the total fluxes (neoclassical and turbulent) and the particle and heat sources. This iterative process is repeated until a flux equilibrium is reached. Subsequently, the resulting profiles are used as input for the GENE-3D code, and the procedure is repeated until TANGO reaches a flux equilibrium once again. A notable distinction is that GENE-3D incorporates the radial electric field in the calculations instead of its shear.
\begin{figure}[!t]
\begin{center}
\begin{tabular} {c}
\includegraphics[width=0.48\textwidth]{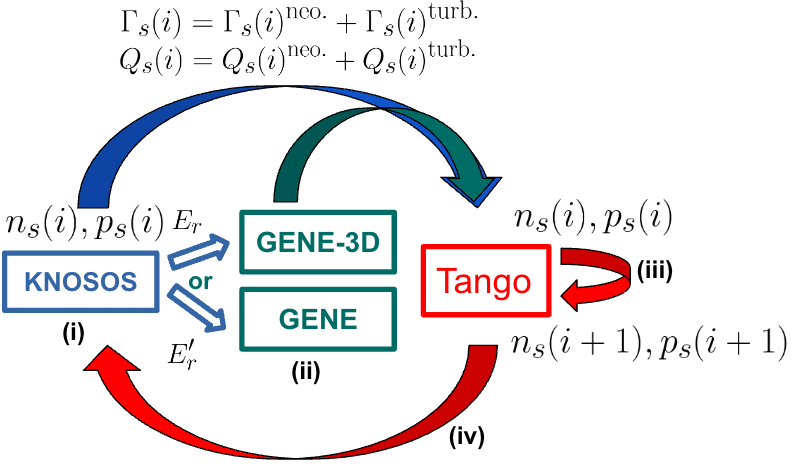} 
\par
\end{tabular}
\end{center}
\vspace{-0.7cm}
\caption{Workflow diagram for the employed theoretical framework:
(i) Given an initial set of density and pressure profiles for each species ($n_s(i), p_s(i)$), KNOSOS evaluates the neoclassical particle and heat fluxes $(\Gamma_s(i)^{\rm neo.}, Q_s(i)^{\rm neo.})$, and the background radial electric field $E_r$.
(ii) GENE computes the turbulent particle and heat transport $(\Gamma_s(i)^{\rm turb.}, Q_s(i)^{\rm turb.})$ using these profiles (including $E_r'$). (iii) Tango evaluates new plasma profiles that are consistent with both the overall (neoclassical and turbulent) fluxes as well as the particle and heat sources. (iv) The process is repeated until a flux equilibrium is reached. The obtained profiles serve as input for GENE-3D, replacing GENE, and the iterative process continues until TANGO achieves a state of flux equilibrium once again.}
\label{fig:fig0}
\end{figure}

This theoretical framework is uniquely equipped to capture a broad range of 3D effects, such as globally radial ones, including internal transport barriers~[\onlinecite{garbet_2002, Strugarek_2013, disiena_2021}], turbulence spreading~[\onlinecite{Garbet_1994, Diamond_PPCF2005, lin_2004}], and transport avalanches~[\onlinecite{pradalier_2010, mcmillan_2010, sarazin_2000}]. Additionally, it can handle magnetic field-line dependencies~[\onlinecite{Xanthopoulos_2014}, \onlinecite{disiena_2020}], neoclassical effects~[\onlinecite{Watanabe_2011, Riemann_2016, pavlos_2020}], as well as external particle/heat sources. The inter-code coupling approach takes advantage of the time scale disparity between turbulence and transport phenomena, and each code is run only on its natural time scale~[\onlinecite{canday_2019}, \onlinecite{barnes_2010}], leading to a drastic reduction in computational cost for evolving plasma profiles up to the energy confinement time~[\onlinecite{gene_tango_paper}]. Furthermore, the influence of 3D effects can be easily assessed in this framework by comparing the solution obtained using the local flux-tube version  with that obtained using GENE-3D.

\end{section}


\begin{section} {Simulation setup \label{setup}}

As discussed above, we aim at investigating the confinement properties of two recently found stellarator configurations, HSK and QSTK, within the theoretical framework presented in the previous section. 
HSK is a quasi-helical stellarator optimized for a large critical gradient for the onset of ITG modes (in terms of $a/L_{T,i}$)~[\onlinecite{robergclark2022reduction}]. It exhibits excellent quasisymmetry, leading to low neoclassical transport and negligible losses of alpha particles. However, it possesses a large vacuum magnetic well ({\em i.e.,} is unstable to MHD modes). As indicated in Table ~\ref{tab:stell}, with a number of field  periods of $n_{f,p}=4$ and an aspect ratio of $A = R/a = 4$ ($R$ as the major radius and $a$ as the minor radius), HSK is a highly compact stellarator.
The design of QSTK is based on the same optimization strategy to significantly reduce ITG turbulence.  However, QSTK combines this aspect with a wider range of desirable attributes for a stellarator reactor concept, including low neoclassical transport, effective alpha particle confinement, MHD stability, and an acceptable level of coil complexity~[\onlinecite{qstk}]. This configuration is characterized by a higher number of field periods, namely $n_{f,p}=6$, and an aspect ratio of $A= R/a = 7.5$.

These two configurations will be compared to HSX~[\onlinecite{hsx}], a quasi-helical stellarator at the University of Wisconsin-Madison, which was not designed for reduced turbulent transport. HSX has an aspect ratio of around $10$, and $n_{f,p}=4$. To ensure comparability and consistency in our study, we will employ all configurations with the same minor radii ($a = 0.75$) and magnetic field strengths on the magnetic axis ($B_0 = 0.85$ T). This choice guarantees the equivalence of the quantity $\rho_{\star} = \rho_i/a$ at the boundary, where $\rho_i$ represents the ion Larmor radius defined as $\rho_i = v_i / \Omega_i$. In this context, $v_i$ denotes the ion thermal velocity given by $v_i = \sqrt{T_i / m_i}$, and $\Omega_i$ represents the ion gyrofrequency expressed as $\Omega_i = (q_i B_0) / (m_i c)$. Here, $T_i$ denotes the ion temperature, $m_i$ stands for the ion mass, $q_i$ represents the ion charge, and $c$ denotes the speed of light. Furthermore, to isolate the turbulence optimization achieved by HSK and QSTK in comparison to HSX, we will exclude KNOSOS from the TANGO loop. 

\begin{table}[!h]
\centering
\resizebox{\columnwidth}{!}{%
\begin{tabular}{|c|c|c|}
\hline
\textbf{Stellarator} & \textbf{Aspect Ratio ($A$)} & \textbf{Field Periods ($n_{f,p})$} \\
\hline
HSK & 4 & 4 \\
\hline
QSTK & 7.5 & 6 \\
\hline
HSX & 10 & 4 \\
\hline
\end{tabular}%
}
\caption{Comparison of stellarator device specifications.}
\label{tab:stell}

\end{table}

The simulations are conducted using a single ion species, namely hydrogen, and the electrons are taken to be adiabatic, in line with the work that led to the design of HSK and QSTK. This implies that only ion heat transport is captured, while electron heat transport and the transport of particles and impurities are neglected. The density profile (remaining constant in time) adopted for this study, as shown in Fig.~\ref{fig:fig1}, is based on W7-X electron-heated discharges. This profile has low normalized density gradient values $a/L_n$ across the entire radial domain. Moreover, we maintain $T_e=T_i$ after each TANGO iteration, and the boundary condition for $T_i$ is taken to be $0.45$ keV at $\rho_{\rm tor} = 0.8$ for all configurations, where $\rho_{\rm tor}$ is the square-root of the toroidal flux $\Phi$ normalized by its value at the last closed flux surface, denoted as $\rho_{\rm tor}=\sqrt{\Phi/\Phi_{\rm LCFS}}$. 
The selection of temperature boundary values in this study is again based on W7-X electron-heated discharges.
\begin{figure}[!t]
\begin{center}
\begin{tabular} {c}
\includegraphics[width=0.5\textwidth]{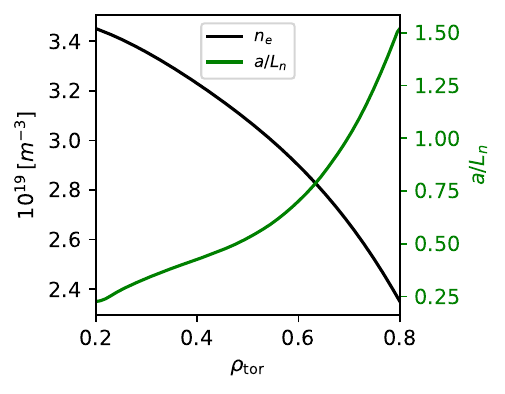} 
\par
\end{tabular}
\end{center}
\vspace{-0.7cm}
\caption{The fixed in time density profile and the corresponding normalized density gradient $a/L_n$ as a function of $\rho_{\rm tor}$ for the simulations utilized in this work.}
\label{fig:fig1}
\end{figure}

In the present work, TANGO is used to solve the ion pressure equation using $27$ radial grid points. At each iteration, we conduct six GENE flux-tube simulations for the magnetic field line defined by $\alpha = q \theta^{\star} - \phi = 0$, covering a range of $\rho_{\rm tor}$ from $0.3$ to $0.8$ with a radial grid spacing of $0.1$. Here, $q$ represents the safety factor, $\theta^{\star}$ is the PEST poloidal angle, $\phi$ is the toroidal angle. The GENE flux-tube simulations use a resolution of $(32, 12, 48, 16, 4)$ in $(k_x,k_y,z,v_{\parallel},\mu)$, where $k_x$ is the radial wavenumber, $k_y$ is the binormal wavenumber, $z$ is the coordinate along the magnetic field line, $v_{\parallel}$ is the parallel velocity, and $\mu$ is the magnetic moment. We set the radial box size to approximately $l_x \approx 64 \, \rho_i$, and the minimum binormal wavenumber used is $k_{y, \rm {min}} \, \rho_i = 0.1$.  In the parallel direction,  the twist-and-shift boundary condition~[\onlinecite{beer_95}] leads to quantization for the radial box size given by $l_x = N / k_{y,min} \hat{s}$, with $\hat{s}$ being the magnetic shear and where $N \ge 1$ is an arbitrary integer ensuring the presence of mode numbers accessed through the parallel boundary condition in the system. However, when the magnetic shear is very small, this may result in a large radial box requiring a high radial resolution. To avoid this, if the shear causes the box size to exceed $20\%$ of the target radial domain box, we set the shear to zero and use parallel periodic boundary conditions in $z$, allowing us to use $l_x = 64 \, \rho_i$. This situation arises throughout the entire radial domain of HSX and extends up to $\rho_{\rm tor} = 0.5$ in HSK and $\rho_{\rm tor} = 0.6$ in QSTK.  Finally, in the parallel velocity dimension, we adopt a box size of $l_{v_{\parallel}} = 2 \sqrt{2} v_i$, while in the magnetic moment dimension, we utilize a box size of $l_{\mu} = 3 \, T_i / B_0$. 
The simulations were conducted using the GPU version of GENE~[\onlinecite{gene_gpu}] on the GPU partition of the Marconi supercomputer (Marconi100) at CINECA. The computational resources allocated for the simulations consisted of $2$ nodes, with each node running for a duration of $2$ minutes per iteration. It is important to emphasize that the chosen resolution aimed to provide a sufficiently fast response rather than an exact flux-tube solution. This rapid response is intended to serve as an initial condition for subsequent GENE-3D simulations, which are computationally more demanding. The use of these initial conditions helps to reduce the number of iterations required in the subsequent simulations, optimizing computational efficiency.

In the GENE-3D simulations, a radial domain ranging from $\rho_{\text{tor}} = 0.2$ to $\rho_{\text{tor}} = 0.8$ is utilized, employing the gradient-driven approach~[\onlinecite{Goerler_JCP2011}]. To maintain consistency with the plasma profiles of TANGO, a Krook-type heat source is incorporated in each iteration. Zero Dirichlet boundary conditions are enforced, and buffer zones covering $5\%$ of the radial domain are employed with a Krook damping operator to mitigate non-physical profile variations near the boundaries. The value of $\rho_{\star}$ at the boundary is approximately $1/\rho_{\star} \approx 300$.
The resolution in the $(\rho_{\text{tor}}, y, z, v_{\parallel}, \mu)$ space is set to $(192, 256, 128, 32, 8)$. Here, the $y$ coordinate represents the direction along the binormal and takes advantage of the periodic-symmetric nature of the stellarators. Specifically, only one-fourth of the toroidal domain is utilized for HSX and HSK, while one-sixth is employed for QSTK. The simulations were performed on 64 nodes of the Marconi CPU Skylake partition at CINECA, with each iteration running for approximately $4$ hours.

\end{section}


\begin{section}{Numerical results \label{results}}

\begin{subsection}{Steady-State Plasma Profiles}

In a first step, we now apply the GENE-3D/TANGO integrated model to investigate the behavior of the three configurations when subjected to an external ion heating source of $5$ MW. The heating power is applied in the region from $\rho_{\text{tor}} = 0.24$, just outside the GENE-3D left buffer region, up to $\rho_{\text{tor}} = 0.29$. Assuming no radiation losses, in flux equilibrium, there will be a constant radial heat flux of 5 MW across each flux surface outside of $\rho_{\text{tor}} = 0.29$. Therefore, we present the results exclusively from the range starting at $\rho_{\text{tor}} = 0.3$ and ending at the boundary at $\rho_{\text{tor}} = 0.8$.
To obtain the steady-state ion temperature profiles, an iterative approach is employed. Initially, an estimate of the profiles is provided, and GENE/TANGO is run for approximately $20$ iterations. These profiles are then utilized as inputs for the GENE-3D/TANGO simulations, which achieve convergence in less than $10$ iterations. This iterative procedure significantly reduces the computational cost of the GENE-3D/TANGO simulations, halving the number of iterations required for convergence.

\begin{figure}[!t]
\begin{center}
\begin{tabular} {c}
\includegraphics[width=0.47\textwidth]{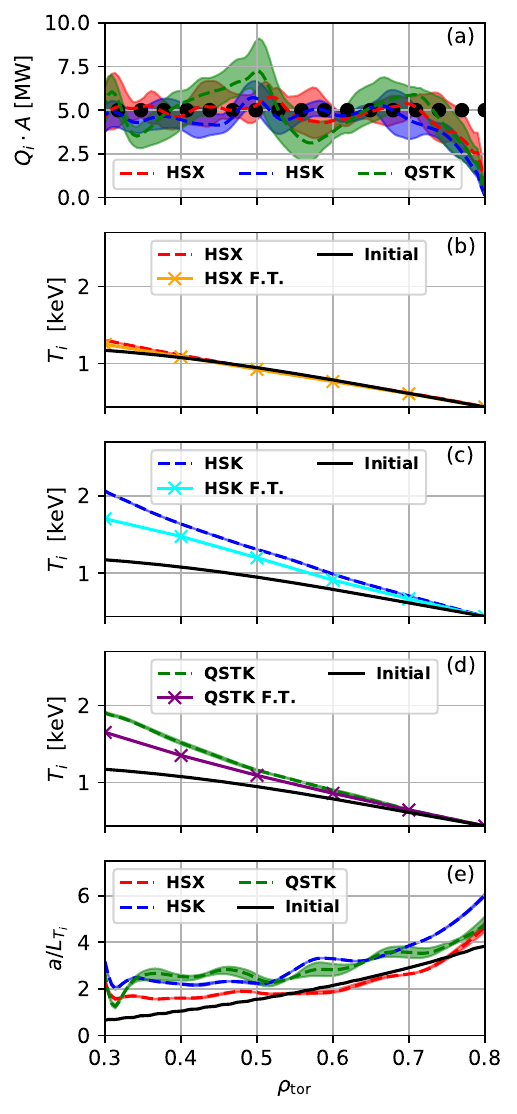} 
\par
\end{tabular}
\end{center}
\vspace{-0.7cm}
\caption{Simulation results using the GENE-3D/TANGO integrated model. 
(a) Ion heat source profiles with a $5$ MW input for HSX (in red), HSK (in blue), and QSTK (in green) as a function of $\rho_{\rm tor}$. The shaded areas in the plot indicate the fluctuations of the heat fluxes over the last five iterations and the circles depict the volume integral of the heat source.
The resulting ion temperature profiles for HSX (b), HSK (c), and QSTK (d) configurations are shown, with the initial profiles indicated by solid black lines and the flux-tube (F.T.) results represented by crosses. (e) The corresponding normalized temperature gradients $a/L_{T,i}$ of the global simulations  for all the configurations.}
\label{fig:fig2}
\end{figure}

The simulation results are presented in Figure~\ref{fig:fig2} as a function of $\rho_{\text{tor}}$. The shaded regions in Figure~\ref{fig:fig2} represent the fluctuations over the last five iterations. The numerically simulated heat flow over the final five iterations agrees well with the value set by the ion heat source, as shown in Figure~\ref{fig:fig2}(a). Figures~\ref{fig:fig2}(b), (c), and (d) depict the ion temperature profiles obtained for HSX, HSK, and QSTK, respectively. In the case of HSX, it is remarkable that the initial guess for the profile (indicated by solid lines) closely aligns with the final flux-tube (F.T.) results (marked by crosses) as well as the global result (represented by dashed lines). The ion temperature at $\rho_{\text{tor}} = 0.3$ is approximately $1.3$ keV. Moving on to HSK, we observe a larger deviation between the initial condition and the final flux-tube result. However, the global simulation closely resembles the flux-tube result. The resulting ion temperature at $\rho_{\text{tor}} = 0.3$ is higher than that of HSX, with $T_i$ close to $2.1$ keV in the global result.
Comparatively, the QSTK results exhibit similarities to the HSK case, particularly the close agreement between the flux-tube and global results, and a higher peak in $T_i$ at $\rho_{\text{tor}} = 0.3$ close to $1.9$ keV compared to HSX.
The significantly higher peak temperature profiles of HSK and QSTK compared to HSX can be attributed to the considerably larger normalized temperature gradient $a/L_{T_i}$ across the entire radial domain, as depicted in Figure~\ref{fig:fig2}(e).  The radial average normalized temperature gradient ratios for HSK and QSTK, relative to HSX, are $35\%$ and $42\%$, respectively, while a naive expectation based on aspect ratio would be approximately $150\%$ for HSK and $33\%$ for QSTK.

Using this framework, the influence of 3D effects can be readily evaluated by comparing the results obtained from the local flux-tube approach with those derived from GENE-3D.  However, to obtain accurate flux-tube results, we performed a GENE/TANGO simulation using a higher resolution in GENE. Specifically, we used $(128, 64, 128, 32, 8)$ grid points in $(k_x, k_y, z, v_{\parallel}, \mu)$. The radial box size was set to approximately $l_x \approx 125 \, \rho_i$, and the minimum binormal wavenumber used was in the range $k_{y, \rm {min}} \, \rho_i = [0.03-0.05]$. We used a box size of $l_{v_{\parallel}} = 3 \sqrt{2} v_i$ in the parallel velocity dimension and $l_{\mu} = 9 \, T_i / B_0$ in the magnetic moment dimension. Using the converged profiles of the lower resolution flux-tube simulation as initial conditions, the GENE/TANGO simulation converged in approximately $10$ iterations, with each iteration taking approximately $4$ GPU nodes and a runtime of $4$ hours per iteration.

The resulting ion temperature profiles and their respective normalized gradients are presented in Figure~\ref{fig:fig3} for all the different configurations.
A high level of agreement is observed between the high-resolution flux-tube (F.T. hres) and the global results for the ion temperature profiles, particularly for HSX and HSK (Figures~\ref{fig:fig3}(a) and (c), respectively). When comparing the normalized temperature gradients in Figures~\ref{fig:fig3}(b) and (d) for these configurations, there is a consistent agreement between the flux-tube results and the global simulations. However, there are slight oscillations in the gradients of the global simulations that are not captured in the flux-tube simulations, likely due to the use of a finer radial grid in the global simulations.
The situation is slightly different for QSTK. The local and global results for QSTK show good agreement in the range $\rho_{\rm tor} = [0.5-0.8]$. However, in the inner core region with $\rho_{\rm tor} \approx [0.3-0.5]$, the global profile exhibits higher ion temperatures compared to the corresponding flux-tube profile due to higher normalized gradients observed in the global simulations (Figure~\ref{fig:fig3}(f)).

Nevertheless, these results are in general,  consistent with the findings of Ref.~[\onlinecite{Navarro_NF_2023}], where local and global simulations with adiabatic electrons for W7-X were also compared, showing also a good agreement between local and global results. However, it is important to note that the application of a kinetic electron model may lead to different results under 3-D effects, as demonstrated in Ref.~[\onlinecite{wilms_2023}], so further investigation is necessary in that regard.
\begin{figure}[!t]
\begin{center}
\begin{tabular} {c}
\includegraphics[width=0.5\textwidth]{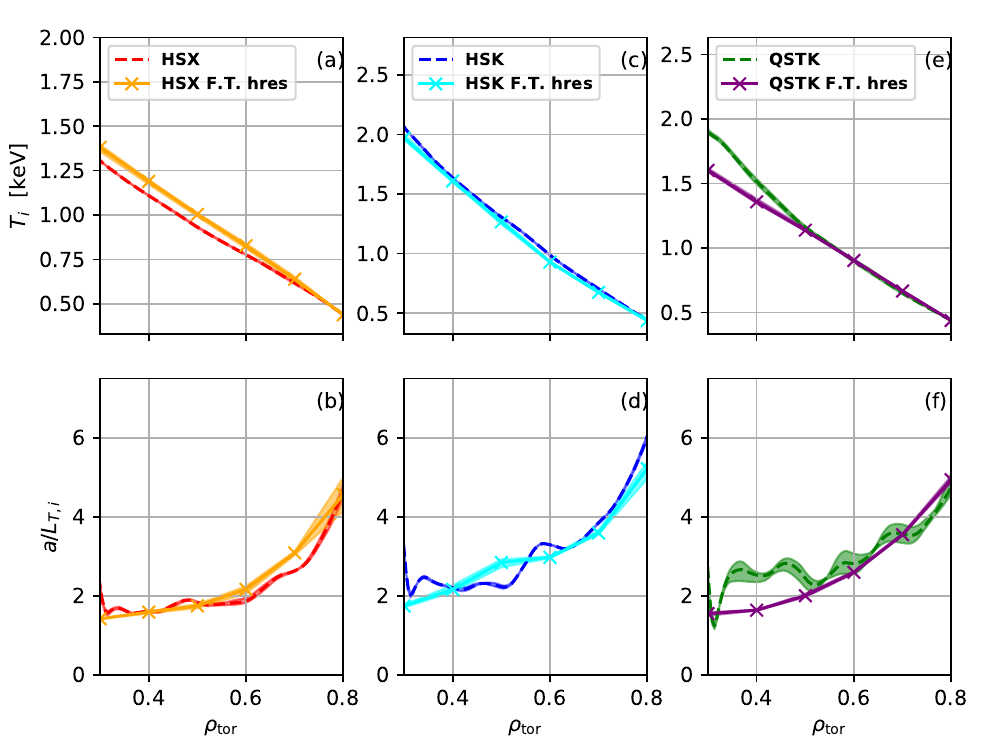} 
\par
\end{tabular}
\end{center}
\vspace{-0.7cm}
\caption{Comparison of ion temperature profiles and normalized gradients between flux-tube  and global simulations for the ion heat source of $5$ MW input power. The resulting ion temperature profiles for HSX (a), HSK (c), and QSTK (e) are shown, with the global results indicated by dashed-lines and the high-resolution flux-tube (F.T. hres) results represented by crosses. The corresponding normalized temperature gradients $a/L_{T,i}$ are shown for HSX (b), HSK (d), and QSTK (f).}
\label{fig:fig3}
\end{figure}
\end{subsection}

\begin{subsection}{Heating Power Sensitivity Analysis}

In this section, we explore the dependence of the results on the heating power.  For this purpose, we conduct two additional sets of simulations with ion input sources of $1$ MW and $10$ MW. The simulation results, presented in Figure~\ref{fig:fig4}, follow the same hierarchical approach as described in the previous section. However, in these simulations, we used the solution obtained from the $5$ MW input power case as the initial condition.

The simulations presented in Figures~\ref{fig:fig4}(a) and (b) demonstrate the successful convergence to the intended input source for the $1$ MW and $10$ MW cases, respectively. Additionally, Figures~\ref{fig:fig4}(c) and (d) display the ion temperature profiles for these power levels. Similarly to the previous scenario with 5 MW input power, it can be observed that the temperature at $\rho_{\rm tor} = 0.3$ is higher for HSK (followed closely by QSTK) compared to HSX.  This temperature difference can be linked to the larger normalized gradients across the entire plasma volume for HSK and QSTK, as illustrated in Figures~\ref{fig:fig4}(e) and (f), in comparison to HSX.
\begin{figure}[!t]
\begin{center}
\begin{tabular} {c}
\includegraphics[width=1.0\columnwidth]{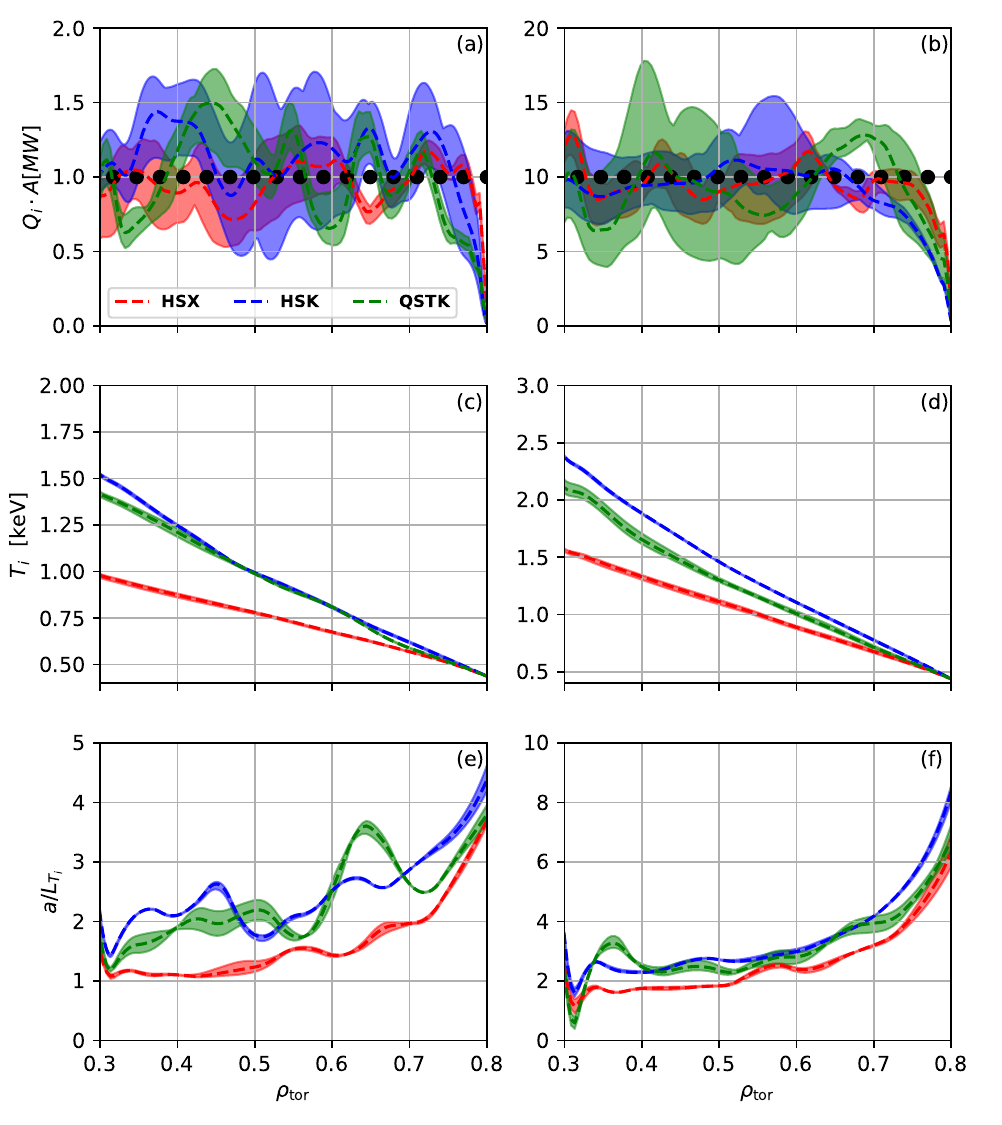} 
\par
\end{tabular}
\end{center}
\vspace{-0.7cm}
\caption{Simulation results comparing HSX (red), HSK (blue), and QSTK (green) configurations as a function of $\rho_{\rm tor}$ for ion heat sources of $1$ MW (left) and $10$ MW (right). Ion heat sources for $1$ MW (a) and  $10$ MW (b), with shaded areas indicating heat flux fluctuations over the last five iterations. The circles depict the volume integral of the heat source. Ion temperature profiles for  $1$ MW (c) and $10$ MW (d), showing the variation with $\rho_{\rm tor}$. Normalized temperature gradients for $1$ MW (e) and $10$ MW (f), for all configurations.
}
\label{fig:fig4}
\end{figure}

We now undertake a more detailed quantitative comparison of the GENE-3D/TANGO results for the ion temperature profile with respect to the ion heating source. The results obtained from GENE-3D simulations, encompassing various ion sources and configurations, are succinctly presented in Figures~\ref{fig:fig6} and~\ref{fig:fig6a}. The data reveals a consistent trend: 
HSK consistently exhibits higher ion core temperatures (Figure~\ref{fig:fig6}(a)), thereby yielding higher ion core to edge temperature ratios in comparison to HSX (Figure~\ref{fig:fig6a}).
Notably, QSTK closely follows HSK with slightly lower temperatures. The temperature ratios of HSK and QSTK relative to HSX range between $1.4-1.6$, and this ratio remains relatively constant as the power levels increase (as observed in Figure~\ref{fig:fig6}(b)).
\begin{figure}[!t]
\begin{center}
\begin{tabular} {c}
\includegraphics[width=0.48\textwidth]{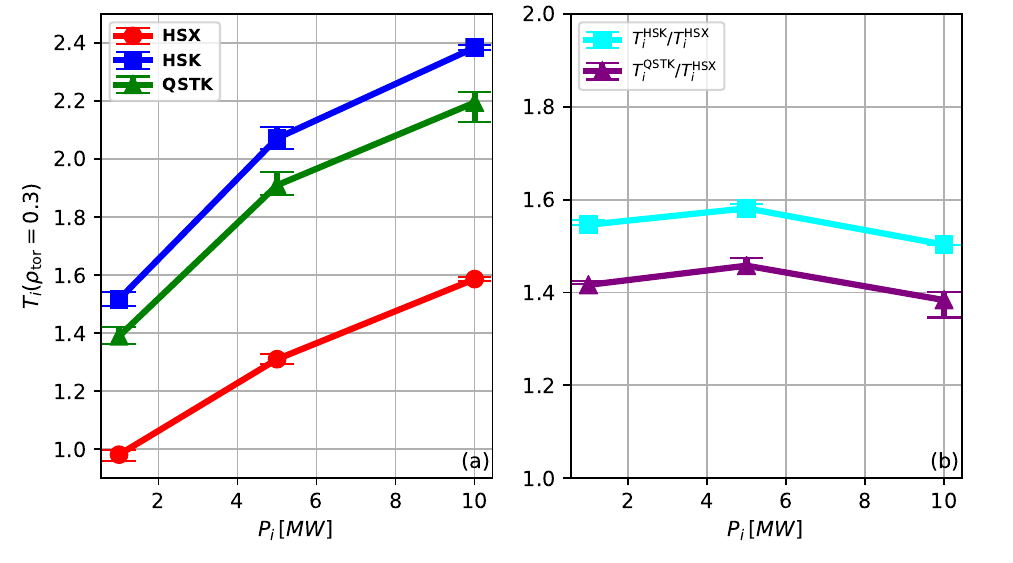} 
\par
\end{tabular}
\end{center}
\vspace{-0.7cm}
\caption {(a) Ion temperature profiles at $\rho_{\rm tor}=0.3$ as a function of the ion heat source $P_i$ for HSX (red circles), HSK (blue squares), and QSTK (green triangles).(b) Ratio between HSK to HSX (cyan squares) and QSTK to HSX (purple triangles) as obtained from the GENE-3D/TANGO simulations.
}
\label{fig:fig6}
\end{figure}
\begin{figure}[!t]
\begin{center}
\begin{tabular} {c}
\includegraphics[width=0.48\textwidth]{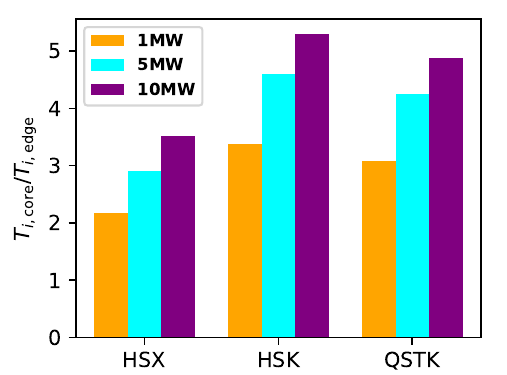} 
\par
\end{tabular}
\end{center}
\vspace{-0.7cm}
\caption {Bar chart of the core ion temperature $T_{i,\rm core}$ ($\rho_{
\rm{tor}} = 0.3$) normalized to the edge $T_{i,\rm edge}$($\rho_{
\rm{tor}} = 0.8$) for all configurations and ion power levels studied in this paper.}
\label{fig:fig6a}
\end{figure}

To investigate whether the increased gradient is primarily influenced by the onset critical gradient of the instability or by the rate of change of the normalized gradient with respect to the normalized heat flux (stiffness), we plotted the normalized temperature gradient $a/L_{T,i}$ against the ion heat flux normalized to GyroBohm units for two representative positions: one in the inner core at $\rho_{\rm tor}=0.4$ and another in the outer core at $\rho_{\rm tor}=0.7$. The corresponding plots are presented in Figures~\ref{fig:fig7}(a) and (b), respectively.

Starting with the outer core position $\rho_{\rm tor}=0.7$, which has a greater influence on determining the final profiles than inner core positions, HSK and QSTK exhibit significantly higher normalized gradients in comparison to HSX. Additionally, by extrapolating the values linearly to the x-axis, the (nonlinear) critical gradient can be estimated. HSK exhibits a critical gradient value of approximately $2.5$, QSTK of $2.2$, followed by HSX with $1.7$. This higher critical gradient is achieved while maintaining a similar level of stiffness ($3.5-4.0$). This observation provides validation for the critical gradient optimization procedure employed in designing the configurations, as they have successfully achieved higher critical gradients while maintaining a similar level of stiffness. However, the situation becomes more intricate when considering other locations. For instance, at $\rho_{\rm tor} = 0.4$, although HSK and QSTK still exhibit higher (nonlinear) critical gradients compared to HSX by approximately a factor of $2$ for HSK, the stiffness at this position varies. HSK demonstrates roughly $3$ times more stiffness than HSX and QSTK. Extrapolating these values to larger gradients suggests that HSK's turbulent transport could potentially surpass that of HSX. This observation implies that as power inputs increase, the disparity in ion temperature ratios between HSK and HSX may further diminish. On the other hand, this larger value of stiffness does not prevent HSK from achieving larger core temperature gradients. Furthermore, QSTK has roughly half the stiffness of HSX at $\rho=0.4$, suggesting further, though as of yet unexplained, benefits of the critical gradient optimization.

\begin{figure}[!t]
\begin{center}
\begin{tabular} {c}
\includegraphics[width=0.48\textwidth]{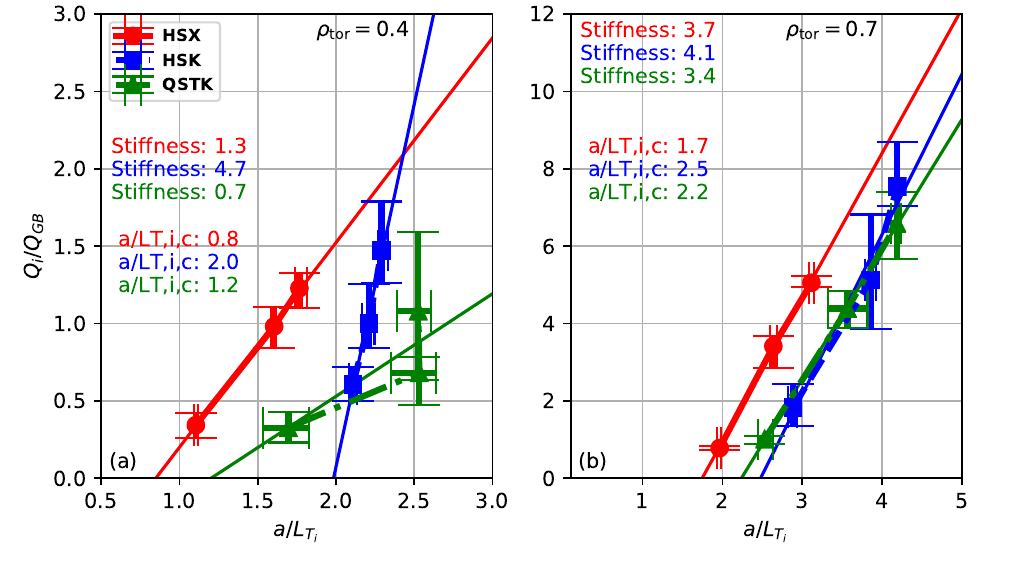} 
\par
\end{tabular}
\end{center}
\vspace{-0.7cm}
\caption{Dependency of the normalized ion heat flux $Q_i/Q_{GB}$ at  $\rho_{\rm tor} = 0.4$ (a)  and at $\rho_{\rm tor} = 0.7$ (b) on the normalized temperature gradient $a/L_{T,i}$ for HSX (red circles), and HSK (blue squares), and QSTK (green triangles) as obtained from the GENE-3D/TANGO simulations. Here, $Q_{GB} = v_i n_i T_i (\rho_i/ a)^2$, with the ion thermal velocity $v_i$ and the ion density $n_i$ evaluated at the corresponding position. The figure also indicates the resulting stiffness and nonlinear (critical) gradient $(a/L_{T,i})_c$ for each configuration, calculated by linearly extrapolating the values to the x-axis (solid lines).
}
\label{fig:fig7}
\end{figure}

\subsection{Confinement Time.}  To assess the overall plasma performance of the different stellarators, we compute the energy confinement time as calculated via the formula
\begin{align}
\tau_{E} = \frac{3}{2} \frac{\int n_i \,T_i \,   d V}{P_H},
\end{align}
where $P_H$ is the heating power and $V$ represents the plasma volume. Here, we have made the assumptions that the electron temperature ($T_e$) is equal to the ion temperature ($T_i$), and that $P_H$ refers only to ion heating. This is to account for the fact that an adiabatic electron model has been used, which does not allow for the determination of electron profiles in a self-consistent manner. In addition it is important to note that, for geometric reasons, the plasma energy will always be dominated by contributions from the outer part of the plasma volume. Although an actual experiment constructed from the HSK or QSTK designs may exhibit significant variation of the edge temperature, due to different confinement in the edge region, such effects cannot be estimated in the present work, and the edge temperatures are therefore set to be equal.  Because of these two facts, we stress that the actual confinement time, as measured in built devices, may differ significantly from the values computed here, and we effectively limit our study to the influence of profile improvements accumulated in the core alone.

Another issue arises in the comparison, stemming from the fact that HSK and QSTK are theoretical stellarator configurations, without defined sizes, and the sensitivity of confinement time to plasma volume makes it somewhat unclear how to compare against an actual experiment like HSX.  The minor radius, for instance, can be freely varied for HSX and QSTK (of course aspect ratio and other dimensionless geometric quantities remain fixed) to produce different results for the confinement time. It can be argued that setting equal volume and magnetic energy density for each configuration provides a reasonable point of comparison for confinement at a similar construction cost, which favors compact stellarators [\onlinecite{Helander_2015}]. Indeed, taking into account increased confinement at fixed volume and the increased core temperature profile, QSTK shows an overall improvement of roughly $80\%$ with respect to the triple product $nT\tau_{E}$ when compared to HSX. Another comparison, at fixed minor radius, is also helpful in the case where a certain minor radius, admitting space for blankets in a reactor, is desired. This comparison favors larger aspect ratios and thus reflects favorably on HSX.

A third way to compare the devices is to try to normalize the confinement time relative to some baseline expectation, for example through a scaling law.  The idea here is to try to separate out the expected dependence on bulk geometric parameters, like the major and minor radius, though this is also challenging to theoretically define in a rigorous way, especially in the complex geometry of a stellarator.  Note that even the well-known ISS04 scaling~[\onlinecite{Yamada_2005}] does not neatly account for the variation among stellarator devices, with rescaling factors required for agreement even among different operational regimes of a single stellarator.  Nevertheless, a simple Ansatz, assuming that the ion temperature profiles (with prescribed edge values) stay relatively close to the linear threshold, leads to the following expression (see Appendix~\ref{appendixa} for more details):

\begin{align}
\tau_E  \propto a^{2 + \beta} R^{1 -\beta} P_H^{\gamma -1}.
\end{align}
Here, the numerical parameters $\beta$, and $\gamma$ are unknown and need to be determined. Interestingly, the Gyro-Bohm scaling presented in Ref.~[\onlinecite{Stroth_nf_2021}] satisfies this very scaling, with $\beta = 0.2$ and $\gamma = 0.4$. In addition, the ISS04 scaling can be considered as a generalization of this expression, using slightly different parameters for the minor radius and major radius. Specifically, one gets $\tau_E\propto a^{2 + \beta_1}R^{1-\beta_2} P_H^{\gamma-1}$, with $\beta_1 = 0.28$, $\beta_2 = 0.36$, and $\gamma = 0.39$.  However, ISS04 scaling cannot be used to evaluate the results obtained here, because of the role of edge temperature is unaccounted for.

In our specific case, we will instead determine appropriate values for the parameters based the simulation data itself, {\em i.e.,} on the analysis of the energy confinement time as a function of the input power ($P_H)$, major radius ($R$), and minor radius ($a$). Regarding the latter, since all simulations were performed with the same minor radius, we assume that we are operating in the limit of $\rho_{\star} \rightarrow 0$, which is supported by the agreement between the local and global simulations. In this limit, the resulting ion temperature $T_i$ does not depend on the minor radius. Therefore, the increase in confinement time is solely due to the change in volume.
\begin{figure}[!t]
\begin{center}
\begin{tabular} {c}
\includegraphics[width=0.5\textwidth]{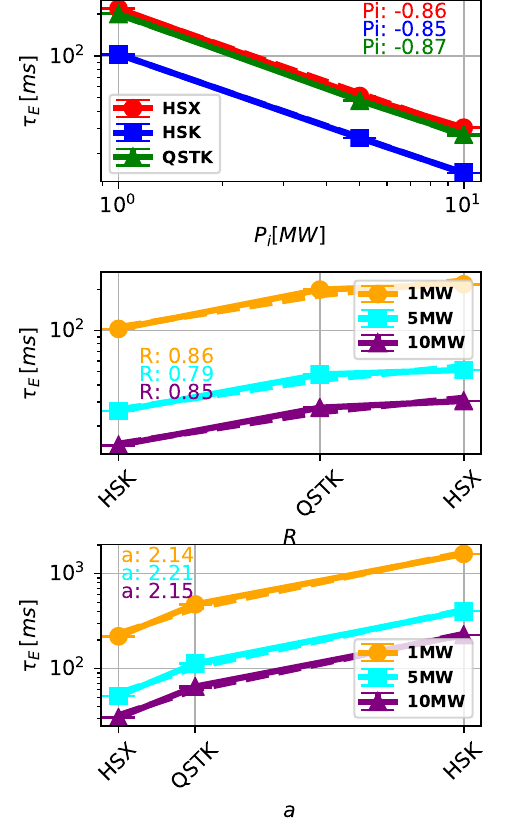} 
\par
\end{tabular}
\end{center}
\vspace{-0.7cm}
\caption{Energy confinement time (in ms) as a function of: (a) Ion heating (in MW), (b) major radius, and (c) minor radius. The dashed lines represent the fitted curves for each respective case, with their corresponding slope values indicated.
}
\label{fig:fig8}
\end{figure}

The results are presented in Figure~\ref{fig:fig8}. In Figure~\ref{fig:fig8}(a), it is observed that all configurations exhibit a power degradation that scales with $P_H^{-0.85}$. This scaling exponent is higher than the values expected from the ISS04 scaling or derived from the Gyro-Bohm scaling laws, which predict a scaling of approximately $P_H^{-0.6}$. It may be that the higher power degradation observed in the simulations may be attributed to the fixed edge ion temperature for all heating powers, while experimental observations indicate an increase in the edge ion temperature with $P_H$. Furthermore, in Figure~\ref{fig:fig8}(b), a scaling law of approximately $R^{0.83}$ (obtained by averaging over all power levels) is observed. Finally, in Figure~\ref{fig:fig8}(c), a scaling with $a^{2.17}$ is also obtained. The observed scaling ($\beta = 0.17$) is relatively close to that assumed in the Gyro-Bohm scaling~[\onlinecite{Stroth_nf_2021}].  We note that the values of the numerical parameters found here, being derived from only three stellarators, would be expected to change somewhat with the inclusion of additional devices.

In summary, based on these simulations, the scaling law for the energy confinement time is found to be as follows:
\begin{align}
\tau_E \propto a^{2.17} R^{0.83} P_H^{-0.85}.
\end{align}
This expression can now be employed to obtain a normalized energy confinement time for each configuration, similar to the definition of the `H factor' in tokamak physics. The resulting plot, showed in Fig. \ref{fig:fig9}, shows only small variation between the three configurations, with QSTK displaying a modest ($\sim 10\%$) improvement relative to the others.  It is however, not obvious how to judge the significance of this improvement, in light of the fact that confinement time depends so weakly on core temperature, when edge temperatures are held fixed, as previously discussed.  Indeed, consider the fact that a device based on QSTK built at equal volume as HSX would, according to our results, only have a $\sim 20\%$ larger confinement time than HSX, despite having a $\sim 50\%$ larger core temperature.  The $\sim 10\%$ of the advantage that remains, upon normalizing the confinement time of such a device according to the above scaling law, shows that the improved confinement of QSTK can be attributed, to similar degrees, to both the more favorable bulk parameters achieved ({\em i.e.,} aspect ratio or compactness) and to additional more detailed shaping of the magnetic geometry resulting from the optimization.  

We emphasize that the answer to the question of whether QSTK, HSK, or future designs based on critical gradient optimization can bring about significantly larger gains in confinement time, relative to baseline expectations accounting for bulk geometric parameters, ultimately depends sensitively on achievable edge temperature, and is not a question that can be definitively answered here.  One clear conclusion is that the optimization might be applied preferentially to outer radial positions to attain the greatest impact on confinement time.

\begin{figure}[!t]
\begin{center}
\begin{tabular} {c}
\includegraphics[width=0.5\textwidth]{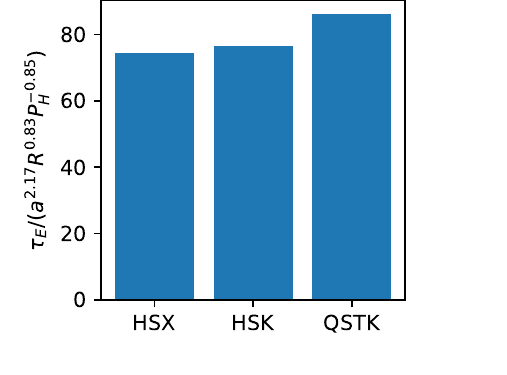} 
\par
\end{tabular}
\end{center}
\vspace{-0.7cm}
\caption{Bar chart of the energy confinement time normalized with respect to the ISS04-like scaling law obtained for the three configurations studied in the present work.}
\label{fig:fig9}
\end{figure}

\end{subsection}


\section{Conclusions \label{conclusions}}

In the present study, we employ a state-of-the-art, first-principles-based theoretical framework to predict the global magnetic confinement properties of two recently found quasi-helical stellarator configurations, HSK and QSTK. Both of them are turbulence optimized in the sense that they exhibit large critical gradients of ITG modes (assuming adiabatic electrons). Here, the existing HSX stellarator (at the University of Wisconsin-Madison) serves as a reference point for judging the achieved quality of the confinement. In line with the studies that led to these two designs, we neglect the effects of kinetic electrons, impurities, magnetic fluctuations, and collisions, thus ensuring self-consistency. Moreover, the edge ion temperature is set equal for all configurations, in absence of a realistic estimate.

The study reveals that HSK and QSTK consistently exhibit significantly higher core-to-edge ion temperature ratios compared to HSX (by up to about 50\%), independently of the applied heating power. This is a reflection of the larger linear critical gradients of ITG modes measured relative to the {\it{minor}} radius of the plasma, which is partly thanks to the significantly smaller aspect ratio of HSK and QSTK, and demonstrates the success of the optimization strategy. Our study also examines profile stiffness, which, on basic theoretical grounds, could lead to deviations from naive expectations based on critical gradients alone.  Here, some variation in stiffness is observed at inner radial locations, but it fortunately does not prove to be a limiting factor regarding the performance.

Secondly, we investigate the energy confinement time, noting that it is fairly insensitive to the central ion temperature when the edge temperature is held fixed, due to the greater weighting of outer plasma volume in the computation of the total stored plasma energy.  It can be concluded that critical gradient optimization could be more influential on confinement time when targeting outer flux surfaces, thus simultaneous increases to core temperature and ion confinement time could be leveraged toward an improved fusion triple product. 
 Above all, we conclude that the evaluation of overall plasma performance will depend crucially on accurate modelling of edge confinement, which is beyond the capabilities of the current tools.  Edge temperature sensitivity precludes the use of empirical scaling laws such as ISS04 to assess plasma performance of HSX, HSK and QSTK using our simulation results.  Comparison of confinement time (and triple product) at equal plasma volume clearly favors the more compact devices, HSK and QSTK, but leaves open questions about what can be considered the cause of that improved performance.  To grapple with this question we further investigate the behavior of ion energy confinement time using a fitted ISS04-like scaling law, in terms of bulk geometric parameters, which shows a small advantage for QSTK in terms of its {\em normalized} confinement time, despite the fact that confinement time, as calculated assuming fixed edge temperatures, averages away much of the effects of the critical gradient optimization.  
 
 The normalized confinement time must be distinguished from the actual confinement time of a machine based on the QSTK or HSK designs, which, depending on size, would differ more significantly from HSX.  Overall, we conclude from the results that the improved ion confinement of both HSK and QSTK relative to HSX can be attributed in part to their lower aspect ratios, although it seems that the advantage exhibited by QSTK shows a similar degree of contribution from additional shaping of the magnetic geometry achieved during optimization.  It is important to also note that even the aspect ratio is a non-trivial geometric parameter; indeed, it is a notoriously difficult quantity to reduce for quasi-symmetric stellarators (in contrast, it is merely a freely adjustable parameter in the equilibrium calculation of tokamaks).  It is also obviously an insufficient target alone for optimizing the turbulence.  Furthermore, it is only one of several properties known to control the critical gradient, due to the tendency of the curvature to roughly follow the inverse major radius, and the possibility therefore remains that future studies employing critical gradient optimization or related methods may lead to configurations that more dramatically exceed expectations from bulk-parameter scaling laws. Integrated strategies could also take into account the possible effect of turbulence on the transport of impurities and the suppression of turbulence as a result of interactions with alpha particles.

\end{section}


{\em Acknowledgements.} The authors would like to thank J.~L.~Velasco and P.~Helander for useful suggestions and comments. Furthermore, we would also like to thank K.~Germaschewski, B.~Allen,~T.~Dannert, and G.~Merlo for their invaluable contributions in developing the GPU version of GENE in the context of the ECP-WDMApp project. This work has been carried out within the frame-work of the EUROfusion Consortium, funded by the European Union via the Euratom Research and Training Programme (Grant Agreement No 101052200 — EURO-fusion). Views and opinions expressed are however those of the author(s) only and do not necessarily reflect those of the European Union or the European Commission. Neither the European Union nor the European Commission can be held responsible for them. Furthermore, numerical simulations were performed at the Marconi Fusion supercomputers at CINECA, Italy.


\appendix 

\section{Simple Energy Confinement Time Scaling Law \label{appendixa}} 

In the following, a simple scaling law for the energy confinement time will be derived, as a reference point to assess the quality of magnetic confinement in various devices. We start by making a few fundamental assumptions:
\begin{itemize}
    \item The edge conditions (ion/electron temperatures and densities) are prescribed. Any dependencies of their values on various device parameters could be added in the context of an extended model, but will be neglected here, in line with the way the simulations are performed.
    \item In the core, which is thought to be in the small $\rho_*$ limit, the ion temperature profiles stay relatively close to the linear threshold of ITG modes, characterized by a critical value of $(R/L_{T_i})_c$. While in practice, this number is expected to vary within moderate bounds, we take it as a constant for our present purposes.
    \item Taking into account that the present investigation as well as the studies leading to the design of HSK and QSTK are based on the assumption of adiabatic electrons, density and electron temperature profiles cannot be determined self-consistently. We therefore assume $T_e=T_i$ and $n_e=n_i \approx$ const.
\end{itemize}
On this basis, we will now estimate the energy confinement time, as defined by
$$ \tau_E = W/P_H \,, $$
where
$$ W = (3/2)\int p\,dV $$
is the stored energy in terms of the plasma pressure $p$, and $P_H$ is the applied heating power.

To compute $W$, let us first introduce the radial coordinate $x=r/a$ and the re-scaled critical gradient $\alpha=(a/L_{T_i})_c$. The ion temperature profile, remaining relatively close to the linear threshold, will vary as $T_i(x)\propto e^{-\alpha x}$, and the electron temperature profile will follow. Consequently, we have
$$ W/V = 3\int_0^1 p(x)\,x\,dx = 3\,p_a\, (e^\alpha-1-\alpha)/\alpha^2 $$
with $p_a=p(1)$ and $(e^\alpha-1-\alpha)/\alpha^2\approx (1/2)\,(1+\alpha/3)$ for $\alpha \approx 0$. In other words, to zeroth order, the stored energy scales with the plasma volume $V$ and the edge pressure $p_a$. The core profile shapes lead to the finite $\alpha$ corrections. In this context, it should be noted that the largest contributions to the overall stored energy come from the near-edge regions, since they account for most of the volume. In the range around and below $\alpha\simeq 1$,  $(e^\alpha-1-\alpha)/\alpha^2$ is proportional -- for a fixed value of $(R/L_{T_i})_c$ -- to a scaling law of the form $(a/R)^\beta$ with $\beta\approx\alpha/3$.

Finally, we also need to address the dependence of $W$ (and possibly $p_a$ in a more comprehensive model) on the heating power $P_H$. It is reasonable to assume that the volume-averaged energy density will increase with increasing $P_H$. This is a reflection of the fact that the profiles tend to steepen (even if only very slightly) under these conditions. Correspondingly, we make the ansatz $W\propto P_H^\gamma$.

In summary, we obtain
$$ \tau_E \propto p_a\,V\,(a/R)^\beta\,P_H^{\gamma-1} $$
or, dropping the dependence on the plasma edge pressure and using $V\propto a^2 R$,
$$ \tau_E \propto a^{2+\beta}\,R^{1-\beta}\,P_H^{\gamma-1}\,. $$
This scaling law is in line with the Gyro-Bohm scaling presented in Ref.~[\onlinecite{Stroth_nf_2021}], which has $\beta = 0.2$ and $\gamma = 0.4$. In addition, it resembles the well known ISS04 scaling~[\onlinecite{Yamada_2005}].



%

\end{document}